\documentclass[prd, aps, superscriptaddress, preprintnumbers, twocolumn, floatfix, nofootinbib]{revtex4}

\usepackage{amsfonts}
\usepackage{amsmath}
\usepackage{amssymb}
\usepackage{bm}
\usepackage{dcolumn}
\usepackage{graphicx}   
\usepackage[latin1]{inputenc}
\usepackage{latexsym}
\usepackage{rotating}
\usepackage{hyperref}
\usepackage{graphicx}
\usepackage{color}

\newcommand\be{\begin{equation}}
\newcommand\ba{\begin{eqnarray}}
\newcommand\ee{\end{equation}}
\newcommand\ea{\end{eqnarray}}

\begin{document}

\title {New Scalar Field Quartessence}

\author{Robert Brandenberger}
\email{rhb@physics.mcgill.ca}
\affiliation{Department of Physics, McGill University, Montr\'{e}al, QC, H3A 2T8, Canada}

\author{Rodrigo R. Cuzinatto}
\email{cuzinatto@hep.physics.mcgill.ca}
\affiliation{Instituto de Ci\^encia e Tecnologia, Universidade Federal de Alfenas,
Rodovia Jos\'e Aur\'elio Vilela, 11999, Cidade Universit\'aria, CEP 37715-400,
Po\c cos de Caldas, MG, Brazil}
\affiliation{Department of Physics, McGill University, Montr\'{e}al, QC, H3A 2T8, Canada}

\author{J\"urg Fr\"ohlich}
\email{juerg@phys.ethz.ch}
\affiliation{Institute of Theoretical Physics, ETH Z\"urich, CH-8093 Z\"urich, Switzerland}

\author{Ryo Namba}
\email{namba@physics.mcgill.ca}
\affiliation{Department of Physics, McGill University, Montr\'{e}al, QC, H3A 2T8, Canada}

\date{\today}

\begin{abstract}

We propose a cosmological scenario involving a scalar field, $\varphi$, that is a source of Dark 
Matter as well as of Dark Energy. Besides $\varphi$, the Lagrangian of the field theory envisaged in our scenario contains a second field $\chi$, for simplicity assumed to be a scalar, too. For fixed values of $\chi$, 
the potential term decays exponentially at large positive values of $\varphi$. While $\varphi$ is not coupled to
Standard Model fields, $\chi$ is assumed to be coupled to them, and the Green functions
of $\chi$ depend on the cosmological redshift in the expanding universe. 
We assume that the term in the Lagrangian coupling $\chi$ to $\varphi$ is such that, at redshifts $z$ larger than some critical redshift $z_c$, $\varphi$ is trapped near $\varphi=0$, and oscillations of $\varphi$ about $\varphi=0$ describing massive scalar particles give rise to
Dark Matter. At redshifts below $z_c$, the field $\varphi$ is no longer trapped near the origin
and starts to ``roll'' towards large field values. A homogenous component of $\varphi$ emerges that acts as Dark Energy. Within over-dense regions, such as galaxies and galaxy clusters, the redshifting of $\chi$ stops, and $\varphi$ therefore remains trapped near $\varphi=0$ as long as $z_c$ is smaller than the redshift when structures on galactic scales decouple from the Hubble flow.
Thus, at the present time, $\varphi$ describes both Dark Energy and Dark Matter.

\end{abstract}

\maketitle

\section{Introduction} 
\label{sec:intro}

The origin and nature of Dark Matter (DM) and of Dark Energy (DE) remain a mystery. There is compelling evidence that DM is particle-like. This would explain the agreement between the amplitude of density fluctuations on large scales and the amplitude of fluctuations in the angular power spectrum of cosmic microwave background anisotropies. Nothing is known about the nature of DE, except that, today, it contributes roughly $70\%$ to the energy density of the Universe, and that its equation of state is close to that of a cosmological constant, namely $w \simeq -1$, where $w = p / \rho$ is the equation of state parameter, with $p$ and $\rho$ denoting pressure and energy density, respectively \cite{Ade:2015xua}; (see, e.g., \cite{DErevs} for reviews on the Dark Energy puzzle).

The most conservative cosmological hypothesis stipulates that DE is a remnant cosmological constant, $\Lambda$, and that DM is a gas of massive particles, called {\it Cold Dark Matter} (CDM) particles, with essentially vanishing pressure. 
The resulting cosmological model is the so-called $\Lambda$CDM Model, which fits the cosmological data on large scales very well. Candidate CDM particles are, e.g., axions, and WIMPS (weakly interacting massive particles), such as the lightest stable supersymmetric partner of Standard-Model (SM) particles, assuming low-energy supersymmetry breaking.\footnote{The energy scale of supersymmetry breaking is chosen such that one can explain, for example, the hierarchy problem in particle physics.} The WIMP paradigm of DM is stringently constrained by experimental data; (see e.g.~the PDG review \cite{Patrignani:2016xqp}, and references therein). 
The hypothesis that a tiny cosmological constant $\Lambda$ is the source of DE raises serious problems: Not only would one want to explain why the cosmological constant is not given by the Planck scale (or the energy scale of supersymmetry breaking in supersymmetric models), energy scales that are about 120 (or 60) orders of magnitude larger than the one corresponding to the observed dark energy density; but one would also want to explain the coincidence that the cosmological constant is becoming important just around the present time.\footnote{See, however, \cite{RHBbr} for a proposal in which a large cosmological constant relaxes towards a dynamical fixed point at which the energy density corresponding to a remnant cosmological constant tracks the matter energy density at late times.
There are several other proposals designed to account for the coincidence problem, such as models with interactions between (dark) matter and DE, (see e.g.~\cite{DErevs} and references therein).} Furthermore, it has recently been pointed out that a bare positive cosmological constant may be incompatible with string theory \cite{swamp}.

It is thus of interest to search for alternative scenarios explaining the origin of the ``dark sector'', i.e., of Dark Matter and Dark Energy.
In particular, one may wish to explore the possibility that DM and DE have a common cause related to degrees of freedom (possibly gravitational ones related to extra dimensions) not taken into account, so far,
which, in our scenario, are described by a single field. This field is assumed to have negligibly weak direct couplings to visible matter and has therefore not been observed directly, so far. Models based on ideas of this kind are sometimes called {\it Unified Dark Matter} (UDM) models; (see e.g.~\cite{UDM} for a review). The most significant challenge facing this approach is to come up with a framework in which a single ``dark component'' can simultaneously yield a phenomenologically realistic, spatially inhomogeneous distribution of DM and a background density of DE that is homogeneous over huge distance scales and is becoming significant at the present time -- rather than only pure DM, at early times in the evolution of the Universe, and pure DE, at late times. Most ``fluid'' models of UDM, as well as various models in which the ``dark component'' is realised by a single field do not meet this last requirement and, hence, are not viable.

An interesting feature of scenarios where the ``dark sector'' is modeled by a new field is that they may suggest extensions of General Relativity connected to invisible extra dimensions and/or novel connections between theories of Dark Matter and Dark Energy, on one hand, and high-energy particle physics, on the other hand. In the past, a number of such approaches have been proposed (see \cite{UDM} for a review). In \cite{CFG}, an extension of the Standard Model has been studied in which a scalar field $\varphi$ with an exponential potential (similar to the one introduced below) appears as a gravitational degree of freedom connected to an
"invisible" discrete extra dimension. Recently, it has been suggested to model DM with a field that has a non-canonical kinetic term and acts as a superfluid on small scales (e.g.,~galactic scales) and as conventional CDM on larger scales \cite{Khoury}. This model can be refined in such a way that the new field might also be a source of DE at late times, \cite{us}. Another (earlier) attempt has been to model DM and DE with a single complex scalar field whose angular component is an axion \cite{Juerg}. Various further proposals have been considered in \cite{Stephon, other}. 

In this paper we describe features of a model of a scalar field, $\varphi$, that might be the common source of Dark Matter \textit{and} Dark Energy. DE is described in terms of a mode of $\varphi$ that becomes dominant after a late-time phase transition (a ``cosmological wetting transition'') and is spatially homogeneous over very large distance scales; DM originates in particle-like oscillations of $\varphi$ near $\varphi=0$, with a mass given by the curvature of an effective potential that confines $\varphi$ to the vicinity of the origin, $\varphi=0$, before the phase transition takes place. This transition is driven by a second scalar field, $\chi$. At  redshifts higher than some critical  redshift $z_c$, effects originating from the coupling of $\varphi$ to $\chi$ yield a contribution to the \textit{effective} potential of $\varphi$ that leads to confining this field to the vicinity of the origin. Below $z_c$, the terms in the effective potential caused by the coupling of $\varphi$ to $\chi$ are so small that $\varphi$ is no longer confined to the vicinity of the origin, and a homogeneous component of $\varphi$ emerges. This ``cosmological wetting transition'' is expected to be continuous \cite{Wetting}.

The scenario studied in this paper shows some resemblances to mass-generation mechanisms -- at sufficiently high redshifts (i.e. background temperatures), the field $\chi$ endows the field $\varphi$ with a positive effective mass (see also \cite{Higgs}) -- and to {\it New Inflation}, where a scalar field is confined to the vicinity of a local maximum of its bare potential by finite-temperature effects \cite{new}. Because of this analogy our scenario is dubbed {\it ``new scalar field quartessence''} \footnote{See \cite{Makler} where the term {\it quartessence} was first introduced.}. The phase transition we have in mind has similarities with the ``wetting transition'' in statistical mechanics; see \cite{Wetting,Pfister} and references given there.

In the following section, Sect. \ref{model}, we describe our scenario in more detail and analyse the resulting dynamics of the cosmological background. In Sect. \ref{sec:formation}, we explain why it makes predictions on curvature fluctuations similar to those of the canonical $\Lambda$CDM model. We conclude the paper with a discussion of our results; see Sect. \ref{conclusion}. Explicit models will be described in a separate paper.

We work in natural units, \mbox{with $\hbar = c = k_B = 1$.} The spatially flat Friedmann-Robertson-Walker-Lema\^{i}tre metric   with scale factor $a(t)$ is taken as the cosmological background. Physical time and present time are denoted by $t$ and $t_0$, respectively, and $z(t)$ stands for the associated cosmological redshift. The reduced Planck mass is denoted by $m_{\rm Pl}$. 

\section{The Scenario} \label{model}

The scenario studied in this paper rests on an effective field theory of two scalar fields, $\varphi$ and $\chi$. The field $\varphi$ will turn out to describe Dark Matter and Dark Energy, while the role of the field $\chi$ is to mediate the ``cosmological wetting transition''. The Lagrangian of these fields contains a potential term given by
\be \label{exppot}
V(\varphi, \chi) \, = \, V_0 g(\varphi, \chi) e^{- \varphi / f} \,  ,
\ee
where $V_0$ fixes the scale of the potential, $f$ is the field value over which the potential decays substantially as a function of $\varphi$ at large, positive values of this field, and the polynomially-bounded dimensionless function $g(\varphi, \chi)$ is chosen in such a way that, after integration over $\chi$, the effective action of $\varphi$, which then depends on the  background Green functions of $\chi$, describes the (cosmological wetting) phase transition discussed below. The kinetic terms of $\varphi$ and $\chi$ are chosen to be quadratic in first derivatives of these fields, as usual. 

It is known that there are stringent constraints on the magnitude of the coupling of a dark-energy field to the degrees of freedom of the Standard Model (SM) of particle physics \cite{Carroll}. In accordance with those constraints we assume that $\varphi$ is \textit{not} directly coupled to fields of the Standard Model.  However, we assume that the modes of $\chi$ get excited in the early universe, possibly, but not necessarily, through direct  interactions with Standard Model fields. 

Integrating over the field $\chi$, we obtain an effective action for $\varphi$ that depends on the Green functions of $\chi$ and hence on the state of $\chi$ at a given time. To elaborate on what we have in mind, roughly speaking, let us imagine that the coupling of $\varphi$ to $\chi$ is described by the term
\be
g(\varphi, \chi) \, = \, V_0^{-1} g^2 \varphi^2 \chi^2 \, .
\ee
After functional integration over $\chi$, the effective action of $\varphi$ contains a potential term which, to leading order in the coupling constant $g$, is proportional to 
\be
\delta V(\varphi) \, \sim \, g^2 \langle :\chi^2: \rangle \varphi^2 e^{- \varphi / f} \,  ,
\ee
where the double colons indicate some normal-ordering of $\chi^2$ that makes the expectation value of this field vanish at zero temperature. We assume that, in the early universe, $\chi$ has been in a state of local thermal equilibrium at some temperature $T$ homogeneous over distance scales corresponding to those which we observe today on large scales. This state has been reached through interactions with modes of Standard Model fields. At high temperature $T$,
\be
\langle :\chi^2: \rangle \, \propto \, T^2 \, .
\ee
Once the state of the field $\chi$ ceases to be in equilibrium with the CMB, assumed to happen at a temperature $T_0$, the expectation value 
\mbox{$\langle :\chi^2: \rangle$} will redshift according to the following law:
Assuming that $T_{0} < m_{\chi}$, the energy density $\rho_{\chi}$ redshifts as \textit{matter}, suggesting that
\be \label{eq5}
\rho_{\chi} \, \sim \, m_{\chi}^2 \langle :\chi^2: \rangle \, \sim \, a(t)^{-3} \,.
\ee
Once non-linearities in the density distribution set in, the local expansion of the Universe stops in over-dense regions, and hence the energy density in such regions stays constant and the redshifting of $\langle :\chi^2: \rangle$ stops, while it continues in regions ``outside of non-linearities''. However, in any region of the Universe, the state of $\chi$ can be characterised by an effective temperature $T$ (typically \textit{not} the temperature of the CMB) that depends on the region.  This effective temperature redshifts in all of space until nonlinear structures appear, but is constant in time inside these nonlinear structures once they have frozen out of the Hubble flow, and is hence $\sim$ constant in time inside of galaxies and clusters of galaxies. This is a crucial feature we will use in our analysis. But we do \textit{not} have to assume that the state of $\chi$ is close to a state of local thermal equilibrium at \textit{late} times.  In the following when we talk about temperature we will mean this effective temperature.

Taking the Green functions of $\chi$ to depend on the effective temperature $T$ of a local region in the Universe, as discussed above, we may write the resulting effective potential as
\be \label{effpot}
V_{\rm{eff}}(\varphi, T) \, = \, V_{0} \lambda(\varphi, T) e^{-\varphi / f} \, ,
\ee
where $\lambda(\varphi, T)$ is a dimensionless polynomially-bounded function of $\varphi$ and of the Green functions of $\chi$, which depend on the effective temperature $T$. We assume that the function $\lambda(\varphi, T)$ have the following properties\footnote{In a separate publication we will study implementations of these conditions in concrete models.} 

\begin{itemize}

\item At temperatures larger than some critical temperature $T_c$, the minimum of the effective potential is at a value $\varphi_{\rm{min}}$ near $\varphi = 0$.

\item Below the critical temperature $T_c$ the effective potential $V_{\rm{eff}}(\varphi, T)$ does not have any local minima, with $V_{\text{eff}}(\varphi, T) \simeq V_{0}e^{-\varphi/f}$, at large
values of $\varphi$.

\end{itemize}

Assuming these conditions are satisfied, then we claim that a phase transition takes place at a temperature $\sim T_c$, which we expect to be continuous (``second or higher order''). 

Thus, under the conditions described above, the field $\varphi$ is trapped near $\varphi = 0$ at   redshifts $z > z_c$. It  has an effective mass given by
\be \label{massform}
m_{\varphi} \, \simeq \, m_{\varphi}(T) \, = \, 
\left( \frac{\partial^2 V_{\rm{eff}}(\varphi, T)}{\partial \varphi^2}\bigg\vert_{\varphi = \varphi_{\rm{min}}} \right)^{1/2} \, ,
\ee
 where $T = T(z)$.
In the early universe, the field $\varphi$ contributes a sizable fraction of the initial energy density in the form of a gas of particles of mass $m_{\varphi}(T)$, (without any evolving homogeneous field component). These are the dark-matter particles.

At   $z_c$, a phase transition takes place. As discussed in the following section, we would like the  redshift $z_c$ to be be close to the average effective temperature of the state of $\chi$, determined through Eq. (\ref{eq5}), reached at a time in the evolution of the universe close to the present time - long after the time when structures on galactic- and cluster scales become non-linear, and hence
\be \label{Trange}
1 \, < \, z_c \, < \, 10 \, .
\ee

After the phase transition, the terms in $V_{\rm{eff}}$ confining $\varphi$ to $\varphi = \varphi_{\rm{min}}$ disappear, and a homogeneous component, $\varphi_0(t)$, in the configuration of the field $\varphi$ emerges. 
The phase transition at $T_c$ is expected to be continuous, and hence $\varphi$ will be given by a spatially quasi-homogeneous function depending on time and evolving according to an equation of motion approximately given by the following one:
\be
{\ddot{\varphi}}_0 + 3 H {\dot{\varphi}}_0 \, = \,  V_0 e^{- \varphi / f} f^{-1} \, .
\ee
It is known \cite{Wands} that the dynamics of a scalar field with an exponential potential in the presence of a perfect fluid with equation of state parameter $w$ has two fixed points. The first one is an inflationary fixed point with vanishing energy contribution from the fluid, the second one is a scaling solution in which the energy density of $\varphi$ tracks that of the fluid \cite{Wetterich}. For
\be \label{fcond}
f \, > \, \frac{1}{\sqrt{2}} m_{\rm{pl}}
\ee
the inflationary trajectory dominates. Thus, if we impose the condition that $f$ is much larger than the limit given in (\ref{fcond}), then the late phase in our cosmology will have an equation of state  close enough to $w = -1$ to be consistent with the current bounds \cite{Ade:2015xua}.

However, as we describe in Section \ref{sec:formation}, the phase transition does \textit{not} take place inside galaxies and galaxy clusters, because, as discussed above, the energy density in these regions remains constant and hence the Green functions of $\chi$ stop to redshift at an effective temperature \textit{larger} than $T_c$. A necessary condition for this to hold is that the phase transition in empty regions of the universe take place \textit{after} fluctuations on galactic scales have become non-linear,  i.e. $|\delta \rho / \rho| > 1$. Such regions are gravitationally bound and do not expand with the Hubble flow. Hence, the Green's functions of $\chi$ become constant in time in such regions.  Thus, we require that the redshift of the transition be less than the redshift of re-ionization, when fluctuations on galactic scales become non-linear, $z_{\rm re} \sim 10$; see \cite{Ade:2015xua}.

If we demand that Dark Energy become dominant at approximately the present time we must require that
\be \label{V0}
V_0 \, \sim \, T_0^3 T_{\rm eq}  \, ,
\ee
where $T_0$ is the present temperature of the CMB. This condition is similar to the one required in quintessence models to guarantee that the onset of accelerated expansion be close to the present time. Thus, our model does not solve the ``coincidence problem'' encountered in quintessence models. 

In addition to the normalization condition (\ref{V0}), condition (\ref{fcond}) (on the value of $f$), and the conditions  guaranteeing that the phase transition described earlier take place at a  redshift $z_c$ in the range given by (\ref{Trange}),  a model must be constructed with the properties that the mass mass and the interaction strength of the second scalar field $\chi$ are consistent with the bounds set by non-observation of new physics at accelerators, and that the energy density in $\chi$ is subdominant. 

In our model, the Dark Matter generation may involve a ``non-thermal'' mechanism. This can be seen in the following way: If the $\varphi$-particles were in thermal equilibrium in the early universe and were to freeze out at some temperature $T_{\rm FO}$ (larger than the temperature $T_{\rm eq}$ at equal matter and radiation), then, assuming that $m_{\varphi} > T_{\rm FO}$, the contribution of $\varphi$-particles to the total energy density at the freezeout temperature would be given by (see e.g.~\cite{KT})
\be
\Omega_{\varphi}(T_{\rm FO})  \, \sim \, \left( \frac{m_{\varphi}}{T_{\rm FO}} \right)^{5/2} 
e^{- m_{\varphi} / T_{\rm FO}} \, ,
\ee
and this had better be approximately equal to $T_{\rm eq} / T_{\rm FO}$, in order to obtain an appropriate abundance of Dark Matter. Hence, we must impose the condition
\be \label{freeze}
\left( \frac{m_\varphi}{T_{\rm FO}} \right)^{5/2}
e^{- m_{\varphi} / T_{\rm FO}} \, \sim \, \frac{T_{\rm eq}}{T_{\rm FO}} \, .
\ee
A necessary condition for this equation to have a root is that $T_{\rm FO} \gtrsim 1.2 T_{\rm eq}$, and the mass of the $\varphi$-particles would have to satisfy the bound $m_{\varphi} \ge 2.5 T_{\rm FO} \gtrsim 3 T_{\rm eq}$. 
However, this mass is very small which may conflict with the  condition (\ref{massform}), in which case a non-thermal production mechanism would be needed. In this respect, the Dark Matter candidate described here has similarities with the fuzzy Dark Matter candidates \cite{fuzzy}. In the following, however, we will assume that $\varphi$ is generated thermally at an early stage, although this assumption is not crucial for the discussion
in the next section.

\section{Structure Formation}
\label{sec:formation}

A well-known problem affecting unified models of Dark Matter and Dark Energy becomes visible when one studies structure formation. In a perfect fluid description, the speed of sound $c_s^2$ that enters the equations of motion for the evolution of density- and curvature fluctuations (see \cite{MFB} for a comprehensive review) is
\be
c_s^2 \, = \, w \, < \, 0
\ee
in the Dark-Energy phase. This leads to an instability in the growth of density fluctuations in the Dark-Energy phase (see e.g.~\cite{UDM} for a review, and \cite{Rodrigo} for a recent study in a particular model) . In order to be consistent with data on the power spectrum of large-scale structure and the cosmic microwave background anisotropies, the time at the onset of Dark Energy domination must be so close to the present time that the models cannot explain the supernova data \cite{Rodrigo}.
This problem can be avoided in generalised fluid models, where $c_s^2 \neq w$ and where $c_s^2$ is tuned to be close to zero in the Dark-Energy phase.

If we model the dark fluid by a scalar field with canonical kinetic term coherently oscillating over a Hubble patch in the Dark-Matter phase then a new problem arises: In the Dark-Matter
phase
\be
c_s^2 \, = \, 1 \, ,
\ee
and hence clustering of fluctuations on sub-Hubble scales is not possible. Hence, such a field configuration cannot explain the clustering of matter in the universe \footnote{But see \cite{Misao2} for a different point of view.}. This is why an axion field coherently oscillating over Hubble scales may not be a viable Dark-Matter candidate; (see, e.g.,~\cite{Misao} for an early study).

Note that the potential issue for scalar-field Dark Matter only arises if the back-ground state of the Dark-Matter field describes coherent oscillations. If Dark Matter were formed by axions, the state of the axion Dark-Matter fluid is a state of axion particle excitations produced by an early thermal phase of axions, or by axion domain wall and axion-string interactions; (see e.g.~\cite{Marsh:2015xka,DESY} for reviews). Similarly, in the ``fuzzy Dark Matter'' scenario with an ultralight pseudo-scalar field \cite{fuzzy}, the state of matter is not an oscillating scalar field, but one based on excited particle states forming a \textit{superfluid} \cite{Hui}. In this case, the effective speed of sound is $c_s^2 = 0$, and structure formation on small scales is possible.

Our scenario is consistent with structure formation. In the early phase, a gas of $\varphi$-particles is generated and, once in thermal equilibrium before freeze-out, at temperatures $T > T_{\rm FO}$, it acts as a fluid, with $c_s^2 = 0$ at temperatures $T < m_{\varphi}$. For  $z < z_c$, a quasi-homogeneous field configuration develops, but without destroying the gas of 
$\varphi$-particles. Thus, for  $z < z_c$ the state of $\varphi$ exhibits a homogeneous ``condensate'' which acts as a Dark-Energy fluid coexisting with a gas of massive $\varphi$ particles.

Since the mass term of $\varphi$ vanishes for $z < z_c$, a uniform  gas of $\varphi$-particles would no longer behave as Dark Matter. But if the transition redshift $z_c$ is smaller than the redshift when structures on galactic and galaxy cluster  scales have become nonlinear then the phase transition does not take place in overdense regions,  as discussed above. Hence, in such regions, the state of $\varphi$ continues to describe a gas of scalar particles with a mass $m_{\varphi}(T) > 0$ that act as Dark Matter.
We thus require  $z_c < 10$, but higher than the redshift when Dark Energy begins to dominate the matter content of the Universe, i.e. $z_c > 1$.

If we normalise our model in such a way that the contribution of Dark Energy to the energy density of the universe is $70\%$ at the present time then its predictions concerning structure formation are, to a first approximation, the same as in the standard $\Lambda$CDM model. As discussed in the previous section, an interesting feature of our model is that the equation of state parameter $w_{\rm DE}$ of the Dark Energy component decreases in time towards $w_{\rm DE} = -1$; (see below \eqref{fcond}).

\section{Conclusions and Discussion} \label{conclusion}

We have presented a scenario in which a single scalar field $\varphi$ can simultaneously describe
Dark Matter and Dark Energy. The dynamics of $\varphi$ is affected by its coupling to a second field $\chi$. At redshifts above a critical redshift $z_c$, radiative corrections induced by $\chi$ trap $\varphi$ near a minimum of its effective potential located near $\varphi = 0$, (similarly as in a model of inflation where an inflaton field is trapped near a local maximum of the \textit{bare} potential in the ``new inflationary'' scenario). Thus, at early times, the state of $\varphi$ describes a gas of massive particles and behaves as cold Dark Matter. At a  redshift $z_c$, a continuous phase transition (related to the ``wetting transition'' in statistical mechanics) occurs allowing a homogeneous component of 
$\varphi$ to dominate. This component acts as Dark Energy around the present time. In over-dense regions, the transition does not take place, and a gas of $\varphi$-particles prevails. Thus, the inhomogeneous modes of $\varphi$ continue to describe Dark Matter in spatial neighbourhoods of galaxies and galaxy clusters even at times when the  redshift is below $z_c$.

The proposal we have described in this paper, which involves two fields $\varphi$ and $\chi$,
may also be implemented by using a single complex scalar field $\phi$ with a symmetry breaking
potential whose angular variable corresponds to the $\chi$-field of this paper, and the radial component
to $\varphi$. (It is quite obvious that the construction of this paper carries over to such models if the exponential potential
(\ref{exppot}) used in this paper is replaced by a double well potential, as long as the slow-roll condition
is maintained). In this way, the current work may be seen as a new realisation of the ideas 
put forward in \cite{Juerg}.

The predictions of our scenario for structure formation are very similar to those of the standard $\Lambda$CDM model or to quintessence models. The fact that in our scenario the equation of state parameter $w$ approaches a value $w_{\infty}$ quite close to $-1$ from above, as time increases, is a distinctive (but not
unique) feature of our proposal.

\section*{Acknowledgement}
\noindent
 
RRC would like to thank the Department of Physics at McGill University for hospitality and for collaborations.
RN thanks Shigeki Matsumoto for discussions on general aspects of Dark Matter, and RHB
thanks Elisa Ferreira, Guilherme Franzmann, Justin Khoury for collaborations
on related projects. RHB also thanks the Institute for Theoretical Studies at the ETH Zuerich for hospitality. The research at McGill is supported in part by funds from NSERC and from the Canada Research Chair program. RRC is grateful for support from CAPES-Brazil (Grant 88881.119228/2016-01).

\end{document}